\documentclass[a4paper,11pt,twoside,onecolumn]{article}
\usepackage{amsmath}
\usepackage[affil-it]{authblk}
\usepackage[bottom=3cm,top=3cm,left=3cm,right=3cm]{geometry}
\usepackage{epsf}
\usepackage{latexsym}
\usepackage{amsbsy}
\usepackage{amssymb}
\usepackage[T1]{fontenc}
\usepackage{textcomp}
\usepackage{helvet}
\usepackage{ifthen}
\usepackage{color}
\usepackage{graphicx}
\usepackage{eso-pic}

	\newcommand{\be}{\begin{equation}}
	\newcommand{\ee}{\end{equation}}
	\newcommand{\ben}{\begin{equation*}}
	\newcommand{\een}{\end{equation*}}

	\newcommand{\argg}[2]{(#1 , #2)}

\begin{document}
\title{A Computational Method to Calculate the Exact Solution for Acoustic Scattering by Fluid Spheroids}
\author{J. D. Gonzalez, E. F. Lavia, S. Blanc}
\affil{\small Underwater Sound Division. Argentinian Navy Research Office (DIIV) UNIDEF (National Council of Scientific and Technical 
Research - Ministry of Defense). Laprida 555. (1638) Vicente L\'opez. Buenos Aires. Argentina. silblanc@yahoo.com}
\maketitle
\begin{abstract} 
The problem of scattering of harmonic plane acoustic waves by fluid spheroids (prolate and 
oblate) is addressed from an analytical approach. Mathematically, it consists in solving the Helmholtz 
equation in an unbounded domain with Sommerfeld radiation condition at infinity. The domain where propagation 
takes place is characterised by density and sound speed values $\rho_0$ and $c_0$, respectively, while 
$\rho_1$ and $c_1$ are the corresponding density and sound speed values of an immersed object that is 
responsible of the scattered field. Since Helmholtz equation is separable in prolate/oblate spheroidal 
coordinates, its exact solution for the scattered field can be expressed as an expansion on prolate/oblate
spheroidal functions multiplied by coefficients whose values depend upon the boundary conditions verified at 
the medium-immersed fluid obstacle interface. The general case ($c_0 \neq c_1$) is cumbersome because it requires to solve 
successive matrix systems that are ill-conditioned when $c_1/c_0$ is far from unity. In this paper, a 
numerical implementation of the general exact solution that is valid for any range of eccentricity values and 
for $c_0 \neq c_1$, is provided. The high level solver code has been written in the Julia programming language while a 
software package recently released in the literature has been used to compute the spheroidal functions. 
Several limit cases (Dirichlet and Neumann boundary conditions, spheroid tending to sphere) have been 
satisfactorily verified using the implemented code. The corresponding example scripts can be 
downloaded from the authors' web (GitHub) site. The numerical implementation of the exact solution leads to 
results that are in agreement with reported results obtained through approximate solutions for 
far-field and near-field regimes. Additionally, the new code has been used to extend results reported in the literature.
\end{abstract}


\section{Introduction}

The interaction of harmonic plane sound waves with prolate/oblate spheroids has been widely and 
increasingly investigated, mainly for soft and rigid scatterers, along the last six decades in different 
branches of acoustics \cite{Spence,silbiger,burnett}. In particular, underwater acoustics can be considered 
quite proliferous in reported articles on different aspects of this topic due to its relevance in applications 
to fisheries management and marine ecosystem research.

Many developed models provide exact or approximate solutions with harmonic time dependence for the general 
problem of acoustic scattering by spheroids which have been extensively applied to several aquatic organisms 
and objects immersed in the ocean \cite{Furusawa,Prarioetal,jechetal2015}. Both for soft and rigid scatterers, 
computational implementations of approximate solutions have been published for either near or far-field 
conditions.

However, less attention has received the case of penetrable spheroidal scatterers till 1964 when the 
corresponding analytical solution for the oblate case was published for the first time \cite{yeh1964}. Three 
years later numerically computed results of the scattered wave radiation patterns were reported only for a 
particular case of sound speed ratio, $c_0/c_1=1$, being $c_0$ and $c_1$ are the sound speed in the medium and 
the fluid prolate spheroid, respectively \cite{yeh1967}. In 1988 Furusawa \cite{Furusawa} 
numerically computed acoustic backscattering patterns from fluid prolate spheroids for specific cases when the 
sound speed and density ratios verify  $c_0/c_1\approx 1$ and $\rho_0/\rho_1\approx 1$.
Several authors have presented results in the last years \cite{chu_gorska,Tang_Nishimori_Furusawa2009} 
valid for the same range of parameters. More recently, another reference \cite{Kotsis} considers the general 
case $c_0 \neq c_1$ and $ \rho_0 \neq  \rho_1$ but only for spheroids with low eccentricity.

Mathematically, the problem of acoustic scattering of plane waves with harmonic dependence consists in 
solving Helmholtz equation in an unbounded domain with Sommerfeld radiation condition at infinity. The domain 
where propagation takes place is characterised by density and sound speed values $\rho_0$ and $c_0$, 
respectively, while $\rho_1$ and $c_1$ are the corresponding density and sound speed values of an object 
immersed in that medium that is responsible of the scattered field. Different boundary conditions must be 
verified for the medium-object interface according to the type of scatterer considered, namely, soft, rigid and penetrable or fluid objects (i.e. Dirichlet, Neumann and transmission boundary conditions, respectively). The solution of the scalar wave equation strongly depends on the object's shape and exact solutions only exist for a 
few cases. When the object is a sphere, the canonical case, the exact solution is based on an expansion in 
spherical wave functions \cite{anderson1950sound}. For scatterers of spheroidal shape, the exact solution can 
be expanded on prolate/oblate spheroidal wave functions as a result of applying separation of variables 
to Helmholtz equation expressed in spheroidal coordinates \cite{morse-feshbach, skud}.

In this paper a new computer implementation based on the exact solution for fluid spheroids 
valid for any value of eccentricity and arbitrary $c_0$, $c_1$, $\rho_0, \rho_1$, is provided. The 
implementation of this solution was developed using recently available computational codes by Adelman \textit{et al.} 
\cite{Agd2014, adelman2014semi} (which will be called AGD software from now on) together with a high level layer code 
implemented in the Julia programming language \cite{Julia}. 

As it is well known, for many cases of interest (e.g. $ka\gg 1$, being $k$ the wave number and $a$ a typical 
longitudinal dimension of the scatterer) computation of spheroidal wave functions requires precision beyond 
the one provided by hardware floating point numbers \cite{VanBuren02}(currently 64 bits in the consumer 
market). In fact, the AGD code implements floating point arbitrary precision through the use of a specialized 
C++ library. The Julia language is a relatively new free programming language with some features that make it 
especially attractive for the implementation of the spheroidal wave function related code, that is the 
possibility of having floating point arbitrary precision arithmetic and algebra, both features built in {\it 
right out of the box}. On the other hand, its calculation speed compared to other high-level equivalent 
computational environments constitutes a very advantageous feature.

This paper is structured as follows. In Section \ref{solucion_exacta} the analytical exact solution is 
rederived. Basically, it is a refurbished derivation that closely follows the previous work by Yeh 
\cite{yeh1964} but contemplating both cases, prolate/oblate spheroids, and correcting some minor typos. This section 
provides the basis for subsequent formulation and establishes the nomenclature that is used in the whole 
article. Section \ref{algoritmo} explains the algorithmic procedure of the coefficient's calculation.
In Section \ref{num_res} it is verified that results derived from the numerical implemented solution agree with
well known limit cases, namely, the penetrable spheroid behaves like the impenetrable 
ones (Dirichlet and Neumann conditions) and from the geometrical viewpoint, when the spheroid tends to the sphere.
Computations of near-field are also included as well as a comparison with previous published results.
Section \ref{applications} exposes results of interest for some applications in acoustical oceanography, mainly  in fisherires acoustics.
In the illustrated examples the numerical implementation allows for extending 
the applicability ranges of previously published results \cite{jechetal2015,okumura2003acoustic}.

In Section \ref{sintaxis}, the computational implementation of the 
numerical solution under the Julia idiosyncrasy is described as well as a brief explanation on the sintaxis of 
the presented computational code is given.
Finally, the conclusions of the work are summarized in Section 
\ref{conclusions}.

\section{Analytical exact solution}
\label{solucion_exacta}

\begin{figure*}[thb]
	\centering
	\includegraphics[width=0.40\textwidth]{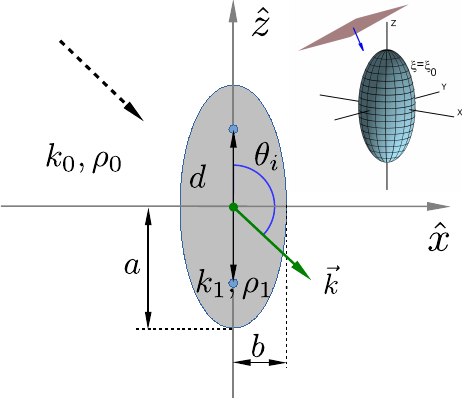}
	\hspace*{10mm}
 	\includegraphics[width=0.40\textwidth]{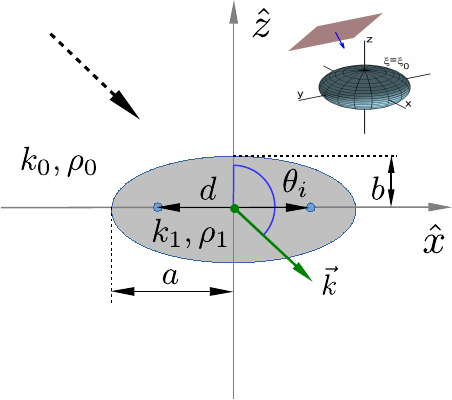}
	\caption{Prolate (left) and oblate (right) coordinate system and related conventions. 
	}
	\label{fig_coord_norm}
\end{figure*}

When acoustic scattering by spheroids is considered, it is convenient to use a spheroidal curvilinear 
coordinate system $(\xi,\eta,\varphi)$ \cite{morse-feshbach}. The relationship between the prolate spheroidal 
coordinates and the Cartesian coordinates, is given by the following transformation \cite{Agd2014}:
\begin{equation*}
\left \{ \begin{matrix} x = \displaystyle \frac{d}{2} \hspace{4pt}[(\xi^{2}-1 ) \hspace{2pt} (1-\eta^{2})]^{1/2} \hspace{4pt} \cos {\varphi}\\
\\
                        y = \displaystyle \frac{d}{2} \hspace{4pt}[(\xi^{2}-1 ) \hspace{2pt} (1-\eta^{2})]^{1/2} \hspace{4pt} \sin {\varphi}\\
\\
                        z = \displaystyle \frac{d}{2}  \hspace{2pt}\xi\hspace{2pt} \eta,     \end{matrix}\right.
\label{transformacionp}
\end{equation*}
where $d$ is the interfocal distance of the ellipse of major semi-axis $a=(d/2) \hspace{2pt} \xi$ and minor 
semi-axis $b=(d/2) \hspace{2pt} (\xi^{2}-1 )^{1/2} $ (See Figure \ref{fig_coord_norm}). The values for the 
three prolate spherical coordinates must lie between the following bounds:
\begin{center}
$1\leq \xi {,} \hspace{10pt}  -1\leq \eta \leq 1{,} \hspace{10pt}   0 \leq \varphi <2\pi$.
\end{center}

The prolate spheroidal coordinate $\xi$-range of variation is directly associated with the shape of the 
spheroid, ranging from $\xi=1$ (corresponding to the segment of length $d$ between both spheroid foci) to 
$\xi$ tending to $\infty$ (corresponding to a sphere of radius $a=b$). The surface of the prolate spheroid 
coincides with the coordinate surface given by equation $\xi= \xi_0$, with $\xi_0= {(1-(b/a)^2)^{-1/2}}$ and 
$d= 2 ({a^2 - b^2})^{1/2}$.

On the other hand, the relationship for oblate spheroidal coordinates is given by
\begin{equation*}
\left \{ \begin{matrix} x = \displaystyle \frac{d}{2} \hspace{4pt}[(\xi^{2}+1 ) \hspace{2pt} (1-\eta^{2})]^{1/2} \hspace{4pt} \cos {\varphi}\\
\\
                        y = \displaystyle \frac{d}{2} \hspace{4pt}[(\xi^{2}+1 ) \hspace{2pt} (1-\eta^{2})]^{1/2} \hspace{4pt} \sin {\varphi}\\
\\
                        z = \displaystyle \frac{d}{2}  \hspace{2pt}\xi\hspace{2pt} \eta,     \end{matrix}\right.
\label{transformaciono}
\end{equation*}
where coordinates ranges are
\begin{center}
$\xi\geq 0$ {,} \hspace{10pt} $-1\leq \eta \leq 1$ {,} \hspace{10pt}  $ 0 \leq \varphi< 2\pi$ {,}
\end{center}
and the surface of the oblate spheroid coincides with the coordinate surface given by equation  $\xi= \xi_0$ where $\xi_0= ((a/b)^2-1)^{-1/2}$.

For harmonic wave fields propagating in an unbounded medium, the acoustic pressure exterior and interior to a 
prolate (oblate) spheroidal surface of a penetrable scatterer immersed in that medium is governed by 
Helmholtz equation 
\begin{equation*}
	\nabla^{2}p + k^{2}p=0 {,}
\end{equation*}
which is separable in prolate (oblate) spheroidal coordinates. Thus, the problem of solving the Helmholtz equation in an unbounded domain with the Sommerfeld radiation at infinity is 
reduced to solving two ordinary differential equations for radial and angular prolate (oblate) spheroidal 
functions, $R(h,\xi)$  and $S(h,\eta)$, respectively \cite{Spence,yeh1964,Agd2014,Abramowitz}, where $h\equiv (d/2) k$ is an adimensional parameter.

At this point is worth noting that the rule $\xi\rightarrow i\xi$  and $h \rightarrow -ih$
provides a direct transformation from the prolate to the oblate case \cite{Agd2014, Abramowitz}. 
Therefore, from now on all mathematical expressions are valid for the fluid prolate spheroid and can be 
easily transformed to the oblate spheroid through the given conversion rule. 

The acoustic pressure corresponding to an incident plane wave with angular frequency 
$\omega$ can be written as
\begin{equation*}
p_i= p_0\exp({i k_0 \hat{k}\cdot\vec{r}}),
\end{equation*}
where $k_0=\omega/c_0$ is the wave number, $\hat{k} = (\sin\theta_i \cos\varphi_i, \sin\theta_i 
\sin\varphi_i, \cos\theta_i)$ is the incidence direction (being $\theta_i,\varphi_i$ the spherical angles of 
incidence) and the amplitude $p_{0}$, a real number.

Without loss of generality, due to the simmetry of revolution around the $z$ axis, we can consider 
$\varphi_i=0$ so that $\hat{k} = (\sin\theta_i , 0 , \cos\theta_i)$. Such incident plane acoustic wave 
on a prolate/oblate spheroid is schematically illustrated in Figure \ref{fig_coord_norm}, including both 
prolate and oblate coordinate systems used in this paper.


The incident pressure can be expanded 
on prolate spheroidal functions \cite{Flammer} so that at an arbitrary field point $\vec{r}= \vec{r}(\xi, 
\eta, \varphi)$ ,
\begin{equation}
	p_{i}(\vec{r}) = 2p_0 \sum_{\substack{m\geq 0 \\ n \geq m}}i^n \frac{\epsilon_m}{N_{mn}} \: 
		S_{mn}\argg{h_0}{\cos(\theta_i)} S_{mn}\argg{h_0}{\eta} \: R_{mn}^{(1)}\argg{h_0}{\xi} \cos(m\varphi), 
	\label{pinc}
\end{equation}
where the subscripts $m,n$ are natural numbers, $S_{mn}$ are the angular spheroidal wave functions, 
$R_{mn}^{(1)}$ are  the radial spheroidal wave functions of first kind, $i$ is the imaginary complex unit 
($i^2=-1$), $\epsilon_m$ is the Neumann factor, defined as $\epsilon_m = 2$  if $m \neq 0$ and $\epsilon_m = 
1$ if $m=0$ and $N_{mn}$ are the norms, given by
\begin{equation*}
	N_{mn} = \int_{-1}^1 S_{mn}^2\argg{h_0}{\eta} \: d\eta.
\end{equation*}
 
The analytical solutions for the scattered and transmitted acoustic pressure, $p_s(\vec{r})$ and $p_t(\vec{r})$, respectively, can be expressed as,
\begin{equation}
 p_s(\vec{r}) =2p_0\sum_{\substack{m\geq 0 \\ n \geq m}} i^n A_{mn} \frac{\epsilon_m}{N_{mn}}S_{mn}\argg{h_0}{\cos(\theta_i)}S_{mn}\argg{h_0}{\eta}R_{mn}^{(3)}\argg{h_0}{\xi}\cos(m\varphi),
\label{pscatt}		
\end{equation}
\begin{equation}
p_t(\vec{r})=2p_0\sum_{\substack{m\geq 0 \\ n \geq m}} i^n B_{mn} \frac{\epsilon_m}{N_{mn}}S_{mn}\argg{h_0}{\cos(\theta_i)}S_{mn}\argg{h_1}{\eta}R_{mn}^{(1)}\argg{h_1}{\xi}\cos(m\varphi).
\label{ptrans} 
\end{equation}

The subscripts 0 and 1 (see Figure \ref{fig_coord_norm}) refers to the surrounding 
medium and to the interior of the prolate spheroidal penetrable obstacle, respectively, and $A_{mn}$ and 
$B_{mn}$ are the unknown expansion coefficients. They are determined applying appropriate boundary conditions 
at the medium-immersed penetrable spheroid interface, that means to 
apply for the continuity of the pressure and the normal component of media particle velocity at the boundary, 
given by $\xi=\xi_0$. In symbols, 
\begin{equation}
	( p_i + p_s)|_{\xi=\xi_0} = p_t|_{\xi=\xi_0}
	\label{bc_presion} 
\end{equation}
\begin{equation}
	\frac{1}{\rho_0} \left. \frac{\partial(p_i+p_s)}{\partial \xi}\right|_{\xi=\xi_0}
	= \frac{1}{\rho_1}\left.\frac{\partial{p_t}}{\partial{\xi}}\right|_{\xi=\xi_0}. 
	\label{bc_deriv}
\end{equation}

In order to compute the $A_{mn}$ and $B_{mn}$ coefficients, it is 
convenient to expand $S_{mn}\argg{h_0}{\eta}$ in terms of $\{S_{m\ell}\argg{h_1}{\eta}\}$
\be
	S_{mn}\argg{h_0}{\eta} = \sum_{\ell=m}^{\infty} \alpha_{n\ell}^{(m)} S_{m\ell}\argg{h_1}{\eta},
	\label{S_expansion}
\ee
where 
\begin{equation}
\alpha_{n\ell}^{(m)}=\frac{\int_{-1}^1 S_{mn}\argg{h_0}{\eta} S_{m\ell}\argg{h_1}{\eta} d\eta }{\int_{-1}^1 S_{m\ell}^2\argg{h_1}{\eta} d\eta}. 
\label{alpha}
\end{equation}
Substituting Equation \eqref{S_expansion} in the LHS of Equations \eqref{bc_presion} and 
\eqref{bc_deriv} and using the orthogonality properties of the family $\left\{S_{mn}(h_1,\eta)\cos(m\varphi): 
m\geq 0, n\geq m \right\}$ lead to a matrix system involving the $A_{mn}$ and $B_{mn}$ coefficients. 

Defining the $Q_{n\ell}^{(m)}$ and $f_{\ell}^{(m)}$ as 
\begin{equation}
 Q_{n\ell}^{(m)} = i^n \frac{1}{N_{mn}} \alpha_{n\ell}^{(m)} \: 
S_{mn}\argg{h_0}{\cos(\theta_i)} \left[ \frac{\rho_1}{\rho_0} \frac{{R}_{mn}^{(3)'}\argg{h_0}{\xi_0} R_{m\ell}^{(1)}\argg{h_1}{\xi_0}}{{R}_{m\ell}^{(1)'}\argg{h_1}{\xi_0}} R_{mn}^{(3)}\argg{h_0}{\xi_0} \right],
\label{QQ}
\end{equation}
\begin{equation}
 f_{\ell}^{(m)}= -\sum_{n=0}^{\infty} i^n \frac{1}{N_{mn}}  
 \alpha_{n\ell}^{(m)}S_{mn}\argg{h_0}{\cos(\theta_i)} \left[ \frac{\rho_1}{\rho_0} \frac{R_{mn}^{(1)'}\argg{h_0}{\xi_0}  R_{m\ell}^{(1)}\argg{h_1}{\xi_0} 
}{R_{m\ell}^{(1)'}\argg{h_1}{\xi_0}} R_{mn}^{(1)}\argg{h_0}{\xi_0}  \right],
\label{ff}
\end{equation}
where the $\xi$ derivative is indicated with a prime (i.e. $ ' \equiv d/d\xi$), it can be shown that the 
$A_{mn}$, with ($n= m, m+1, .. $),  verify  
\begin{equation}
	\sum_{n=m}^{\infty} A_{n}^{(m)} Q_{n\ell}^{(m)}  = f_{\ell}^{(m)}  \;\quad \ell= m, m+1, ... 
	\label{matricial}
\end{equation}
where $m$ is indicated as superscript to emphasize the fact that for each fixed $m$, the left hand side of 
Equation \eqref{matricial} is a product of a row vector $(A_{m}^{(m)},A_{m+1}^{(m)}, \dots, 
A_{n}^{(m)}, \dots)$ and a matrix $Q_{n\ell}^{(m)}$ while the right side is a row indexed by $\ell$.

After having computed the $A_{mn}$, it follows that the other coefficients $B_{mn}$, associated to the 
transmitted field, can also be calculated as  
\begin{equation}
	B_{m\ell} = \sum_{n=m}^{\infty} i^{n-\ell} \frac{N_{m\ell}}{N_{mn}} \: 
	\alpha_{n\ell}^{(m)} \frac{S_{mn}\argg{h_0}{\cos(\theta_i)}}{S_{m{\ell}}\argg{h_0}{\cos(\theta_i)}} \left[ \frac{R_{mn}^{(1)}\argg{h_0}{\xi_0}  + 
	A_{mn}\:R_{mn}^{(3)}\argg{h_0}{\xi_0}}{R_{m{\ell}}^{(1)}\argg{h_1}{\xi_0} } \right] . 
\label{coefB}	
\end{equation}

In the far-field limit it can be shown \cite{yeh1964} that 
$$p_s(r,\theta_s,\varphi_s)  \approx p_0 \frac{e^{ik_0 r}}{r} f_\infty(\theta_s, \varphi_s),$$
where $r,\theta_s, \varphi_s$ are the spherical polar coordinates of the observation point and 
$f_\infty(\theta_s, \varphi_s)$ is the so-called far-field scattering amplitude function which is widespreadly used in different acoustic scattering applications. For the particular case of our spheroid, it is
\begin{equation}
	f_\infty(\theta_s, \varphi_s)= \frac{2}{i k_0}  \sum_{\substack{m\geq 0 \\ n \geq m}} A_{mn} 	\frac{\epsilon_m}{N_{mn}} S_{mn}\argg{h_0}{\cos(\theta_i)}	S_{mn}\argg{h_0}{\cos(\theta_s)} \cos(m\varphi_s).
	\label{finf}
\end{equation}

\section{Algorithmic  procedure}
\label{algoritmo}
The numerical computation of the $A_{mn}$ and $B_{mn}$ coefficients starts with a truncation procedure. The 
series in Equation \eqref{matricial} is truncated at $M$, so that for each value of $m$, there are  $M-m +1$ 
unknown $A^{(m)}_n$, where index $n$ takes natural values in the interval $[m,M]$. Thus, $A^{(m)}_n$ satisfies 
the linear system
\begin{equation}
	\sum_{n=m}^{M}  A_{n}^{(m)} Q_{n \ell}^{(m)} = f_{\ell}^{(m)} \quad \ell= m, m+1, .., M
	\label{matricial_discreto}
\end{equation}
where $A^{(m)}$ and $f^{(m)}$ are row arrays $\in \mathbb{C}^{1\times {(M-m+1)}}$ and $Q^{(m)}$ is a matrix 
$\in \mathbb{C}^{(M-m+1)\times(M-m+1)}$. 

The successive matrix systems obtained as a consequence of truncation, Equation 
\eqref{matricial_discreto}, are 
\begin{itemize}
\item for  $m=0$, 
\begin{equation*}
	\left( A^{(0)}_0, A^{(0)}_1, .. , A^{(0)}_M \right)  \begin{pmatrix}
	Q_{0,0}^{(0)} & Q_{0,1}^{(0)} & \cdots & Q_{0,M}^{(0)} \\
	Q_{1,0}^{(0)} & Q_{1,1}^{(0)} & \cdots & Q_{1,M}^{(0)} \\
	\vdots  & \vdots  & \ddots & \vdots  \\
	Q_{M,0}^{(0)} & Q_{M,1}^{(0)} & \cdots & Q_{M,M}^{(0)} 
	\end{pmatrix}=\left(f^{(0)}_0, f^{(0)}_1, .. , f^{(0)}_M \right) 
\end{equation*}
\item for $m=1$,  
\begin{equation*}
	\left( A^{(1)}_1, A^{(1)}_2, .. , A^{(1)}_M \right)  \begin{pmatrix}
	Q_{1,1}^{(1)} & Q_{1,2}^{(1)} & \cdots & Q_{1,M}^{(1)} \\
	Q_{2,1}^{(1)} & Q_{2,2}^{(1)} & \cdots & Q_{2,M}^{(1)} \\
	\vdots  & \vdots  & \ddots & \vdots  \\
	Q_{M,1}^{(1)} & Q_{M,2}^{(1)} & \cdots & Q_{M,M}^{(1)} 
	\end{pmatrix}  = \left(f^{(1)}_1, f^{(1)}_2,.. , f^{(1)}_M \right)
\end{equation*}
	\item \ldots
	\item for $m=M-1$,
\begin{equation*}
	\left( A^{(M-1)}_{M-1}, A^{M-1}_M \right)  \begin{pmatrix}
	Q_{M-1,M-1}^{(M-1)} & Q_{M-1,M}^{(M-1)} \\
	Q_{M,M-1}^{(M-1)} & Q_{M,M}^{(M-1)} \\ 
	\end{pmatrix} = \left(f^{(M-1)}_{(M-1)}, f^{(M-1)}_M \right)
\end{equation*}
\item and finally, for $m=M$, 
\[
	A^{(M)}_M Q_{M,M}^{(M)} =  f^{(M)}_M.
\]
\end{itemize} 

In summary, the algorithm calculates the $Q^{(m)}$ matrix and $f^{(m)}$ array using the AGD software at each 
step (indexed by $m$) and solves each linear system induced by them at the Julia layer level in order to get 
the  $A^{(m)}_n$ coefficients. Then, the $B^{(m)}_n$ coefficients are easily calculated from Equation 
\eqref{coefB}.
\section{Numerical verifications}
\label{num_res}
\subsection{Limit cases in the far-field zone} 
\label{verifarfield}
Algorithm verification was conducted via evaluation of the exact solution for $|f_\infty|$, obtained through 
the implemented codes for a fluid prolate (oblate) spheroidal scatterer, against known results valid in some 
limit cases (such as spheroid tending to a sphere, non-penetrable Dirichlet and Neumann boundary conditions 
at the medium-spheroid interface).

\begin{figure}[ht]
	\centering
	 \includegraphics[width=0.45\textwidth]{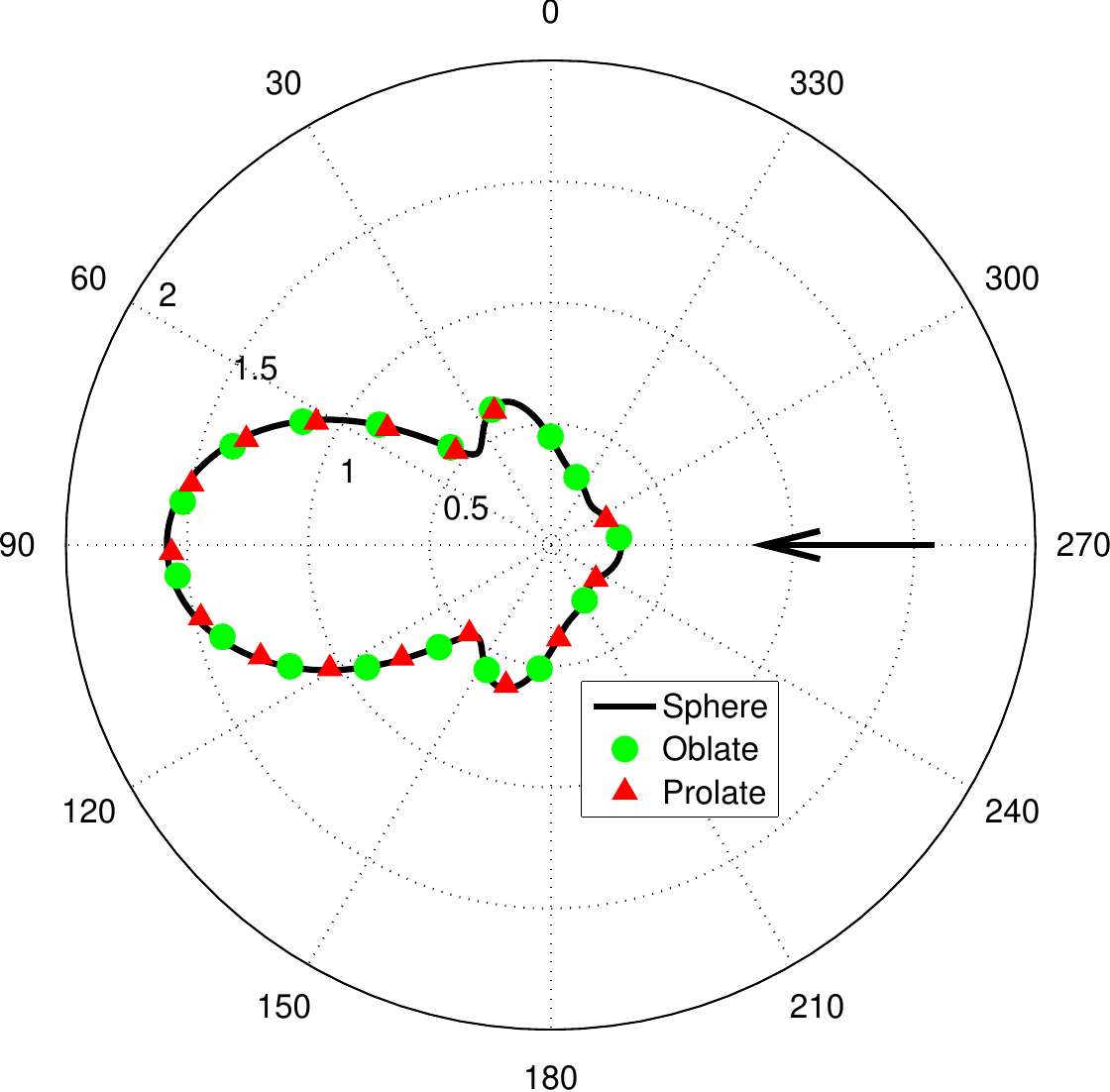}
	\includegraphics[width=0.45\textwidth]{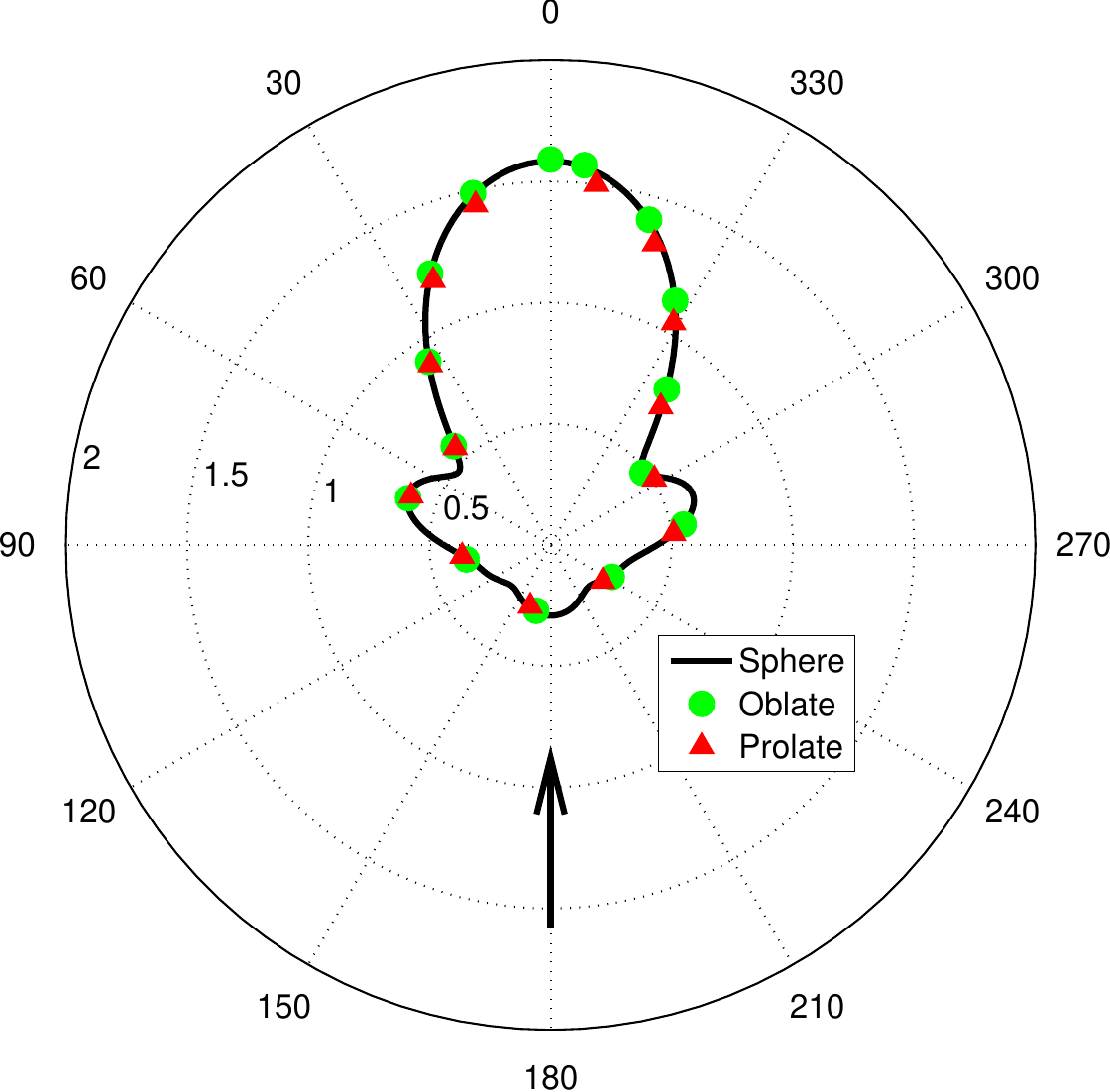}
	\caption{The absolute value of the oblate and prolate far-field scattering amplitude function 
	$|f_\infty|$ tends to the absolute value of the sphere far-field scattering amplitude function when 
	$b\to	a$. The $\hat{x}$ incidence (left) and the $\hat{z}$ incidence (right) are shown.}
	\label{fig_sphere}
\end{figure}

\begin{figure}[ht]
	\centering
	 \includegraphics[width=0.45\textwidth]{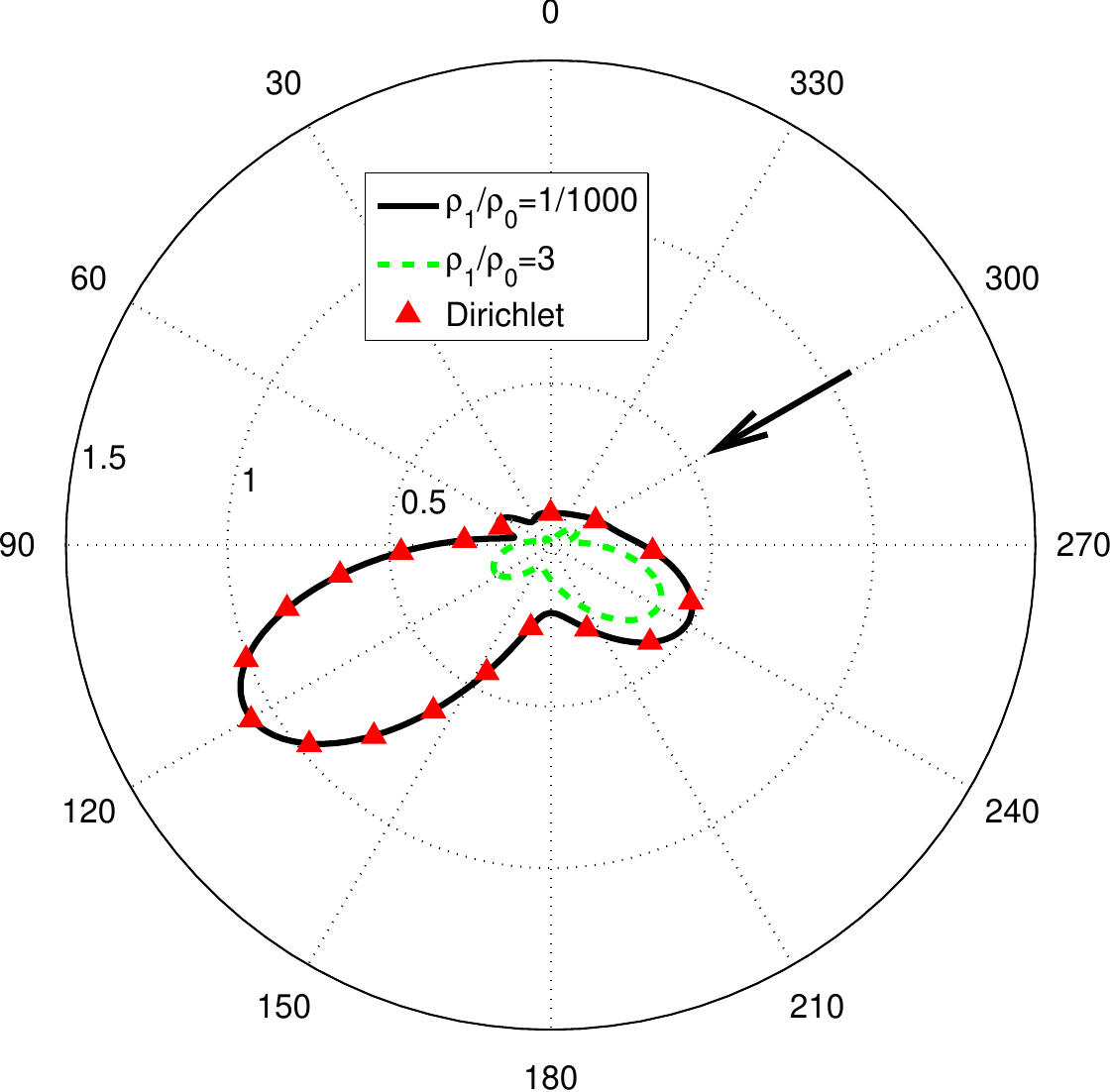} 
 		\includegraphics[width=0.45\textwidth]{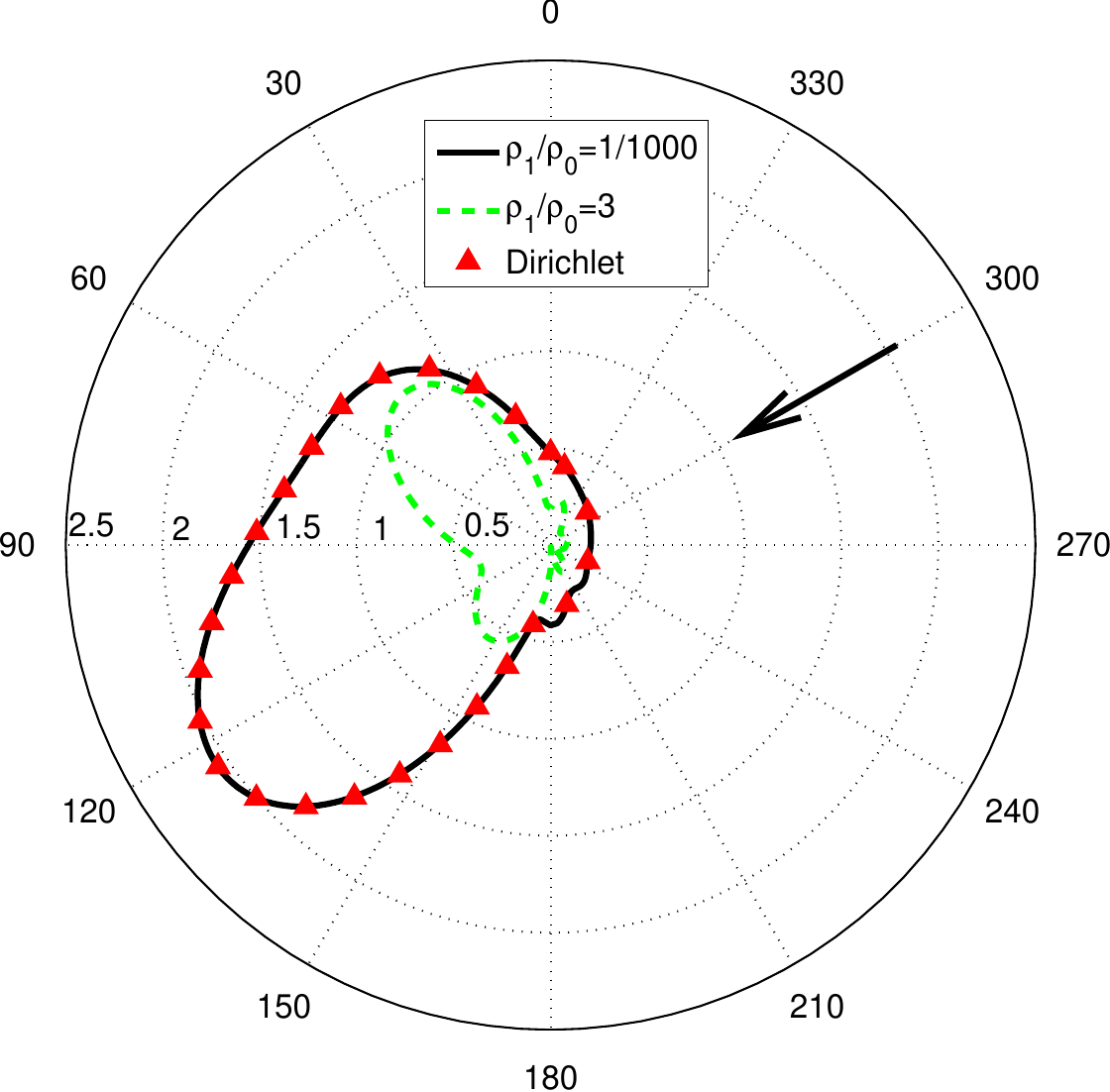}
	\caption{The form function $|f_\infty|$ with two cases of the density ratio and the Dirichlet case.
	Prolate (left) and oblate (right) cases.}
	\label{fig_dirichlet}
\end{figure}

\begin{figure}[bht]
	\centering
  \includegraphics[width=0.45\textwidth]{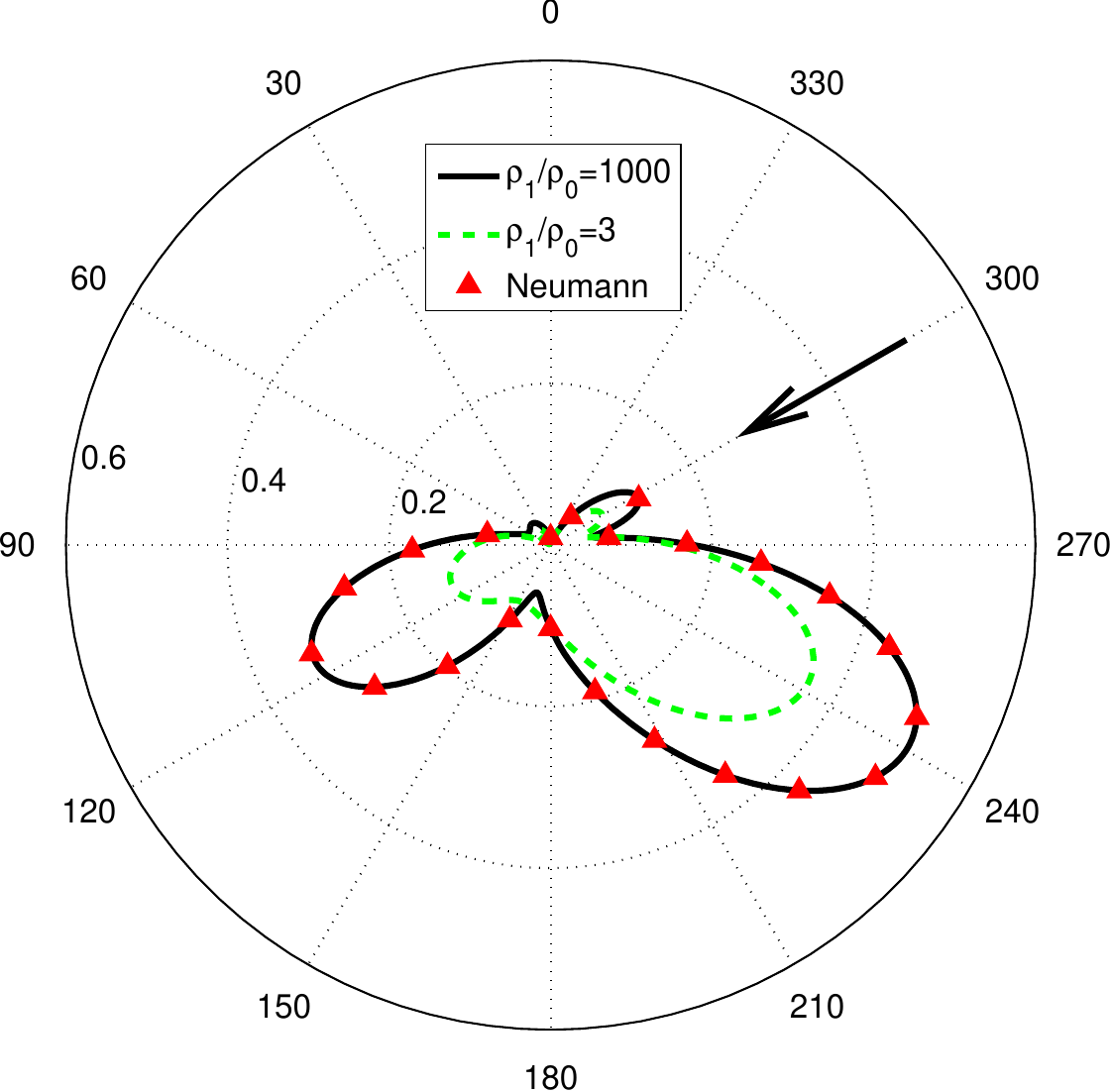}
	\includegraphics[width=0.45\textwidth]{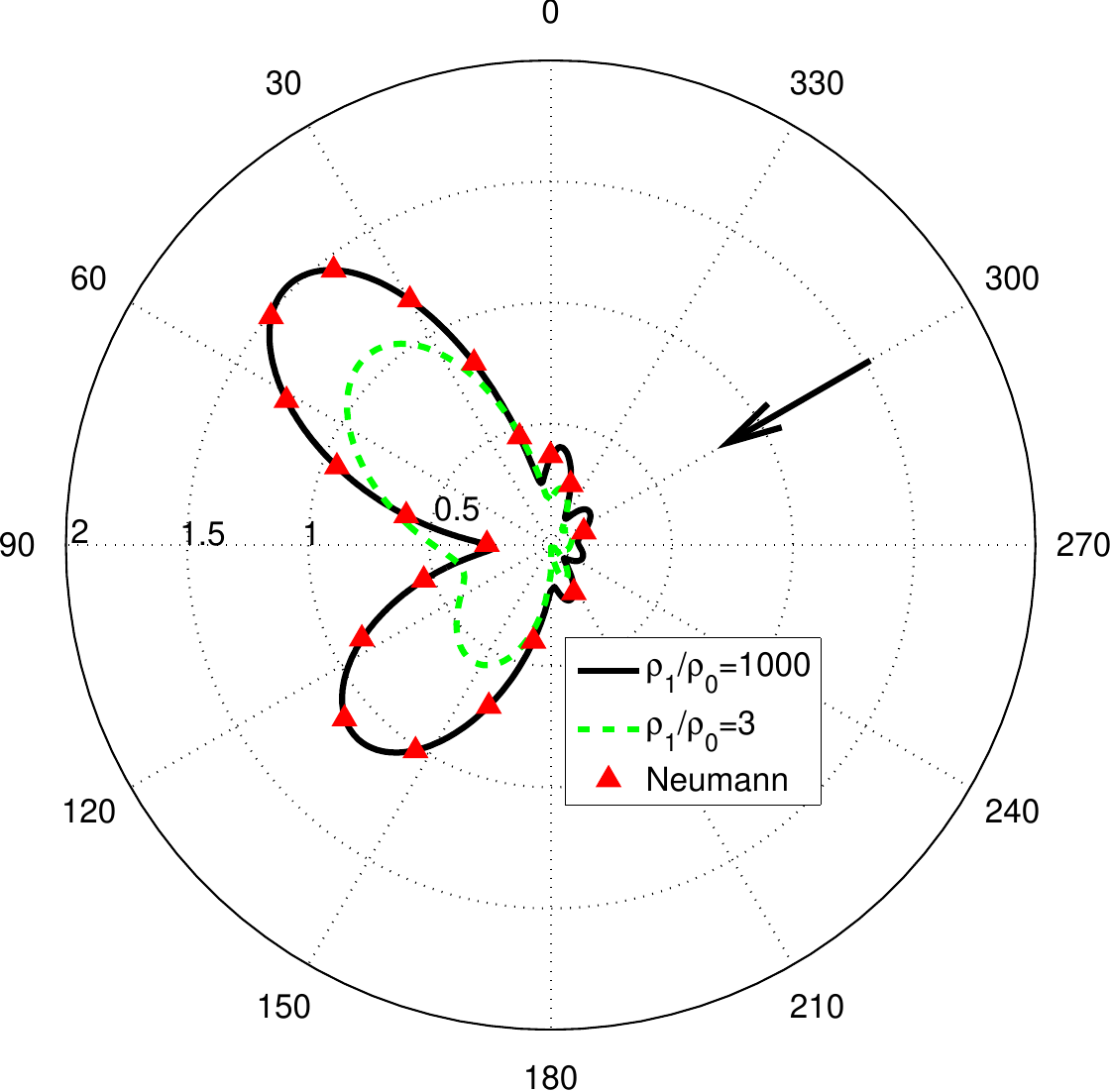}		 	
	\caption{The form function $|f_\infty|$ with two cases of the density ratio and the Neumann case.
	Prolate (left) and oblate (right) cases.}
	\label{fig_neumann}
\end{figure}

A fluid spheroid approaching a sphere, with major semi-axis $a=1$ and minor semi-axis $b=0.99$, is assumed 
in order to compare the results predicted by the implemented codes with the classical solution for plane wave 
acoustic scattering by a fluid sphere \cite{Hickling}. Two different incident directions are considered, 
given by the line that contains both foci of an oblate spheroid (i.e. $\hat{x}$ incidence) 
and of a prolate spheroid (i.e. $\hat{z}$ incidence), in agreement with the coordinate system shown in Figure 
\ref{fig_coord_norm}. The sphere has a reference radius $a=1$. Calculations were made for $k_0=5, k_1=3, 
\rho_1/\rho_0 = 3$. Scattering patterns obtained with the implemented numerical codes for fluid prolate/oblate spheroids 
fits the patterns due to spheres shown in Figure \ref{fig_sphere}.
Furthermore, tests were made to verify that the exact solution implemented for a fluid spheroids tends to 
the known results for impenetrable spheroids, modelling the latter scattererss through extreme density ratios 
in the former ones. For soft spheroid (or what is equivalent, Dirichlet case) \cite{Furusawa, 
Prarioetal, Agd2014}, a density ratio $\rho_1/\rho_0 = 1/1000 $ was assumed, whereas $\rho_1/\rho_0 = 1000 $ 
was used to evaluate the rigid spheroid (equivalent to Neumann case) \cite{Spence, Agd2014}. Computations were 
made for semi-axis values $a=1, b=0.25$ and for an arbitrary incidence angle, $\theta_i=2\pi/3$. Results are 
exhibited in Figures \ref{fig_dirichlet} and \ref{fig_neumann}. It can be observed there that in the limit of 
low density contrast, results for both spheroids tend to reproduce the Dirichlet case (Figure 
\ref{fig_dirichlet}) while in the limit of high density contrast, they tend to reproduce Neumann case (Figure 
\ref{fig_neumann}). Moreover, to illustrate intermediate scatterers' types, patterns for moderate density 
contrasts ($\rho_1/\rho_0 = 3$) are shown.

\subsection{Results for the near-field zone} 

Computation of both, $A_{mn}$ and $B_{mn}$ coefficients is necessary to obtain the scattered and transmitted 
acoustic pressures $p_s(\vec{r})$ and $p_t(\vec{r})$ according to Equations \eqref{pscatt} and  
\eqref{ptrans}.

The total acoustic pressure at a field-point $\vec{r}$ is expressed as
\ben
	p_{total}(\vec{r}) = \left \{ \begin{matrix}  p_{t}(\vec{r}) & \ \ \mbox{ if } \ \  \xi(\vec{r}) \leq 
\xi_0\\ p_i(\vec{r}) + p_s(\vec{r}) & \ \ \mbox{if }  \ \ \xi(\vec{r}) > 
\xi_0\end{matrix}\right.
\een

\begin{figure*}[ht]
	\centering
	\includegraphics[width=0.9\textwidth]{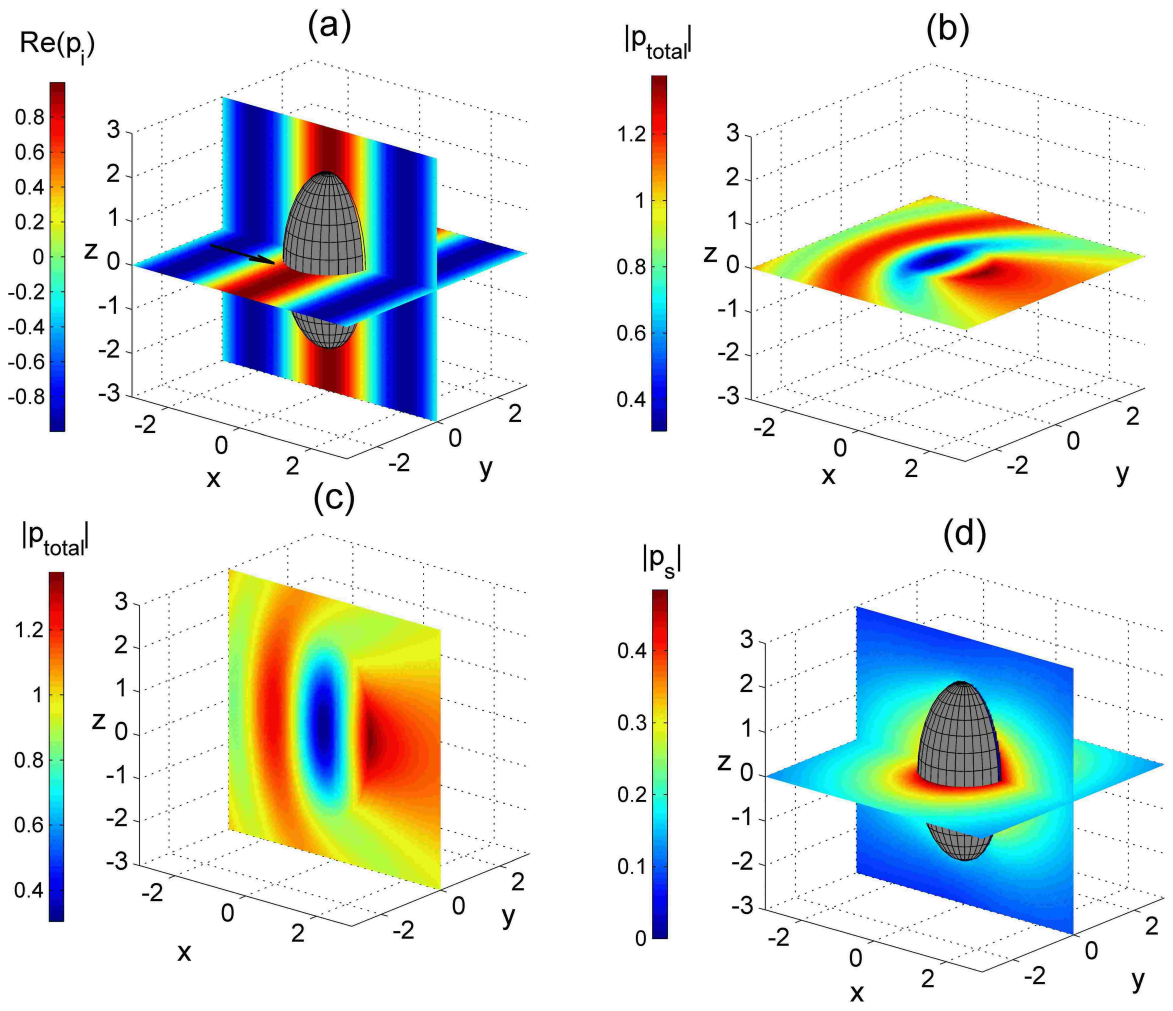}
	\caption{(a) Real part of incident acoustic pressure, $\mathrm{Re}(p_i)$, at the Cartesian 
	planes $z=0$ and $y=0$. (b) Absolute value of total acoustic pressure, $|p_{total}|$, at $z=0$. (c) 
	Absolute value of total acoustic pressure, $|p_{total}|$, at $y=0$. (d) Absolute value of scattered 
	acoustic pressure, $|p_{s}|$, at $z=0$ and $y=0$. Assumed values: $a=2$, $b=1$, density ratio 
	$\rho_1/\rho_0=1.5$, wave numbers $k_1=1$, $k_0=1.5$ and incidence angle $\theta_{i}=\pi/2$. }
	\label{fig:ProNearField}
\end{figure*}

Calculations are held for a prolate spheroid with semiaxes $a=2$, $b =1$, density ratio $\rho_1/\rho_0=1.5$ 
and wave numbers $k_1=1$, $k_0=1.5$. Results for the real part of the incident acoustic pressure $p_i$ and the 
absolute value of total and scattered acoustic pressures are shown in Figure \ref{fig:ProNearField} for the 
Cartesian planes $z=0$ and $y=0$. Computed total acoustic pressure verifies the continuity required by the 
boundary conditions at the interface $\xi=\xi_0$ as it is visually illustrated in Figure 
\ref{fig:ProNearField} $(b)$ and $(c)$. 

\begin{figure*}[bht]
	\centering
	\includegraphics[width=0.47\textwidth]{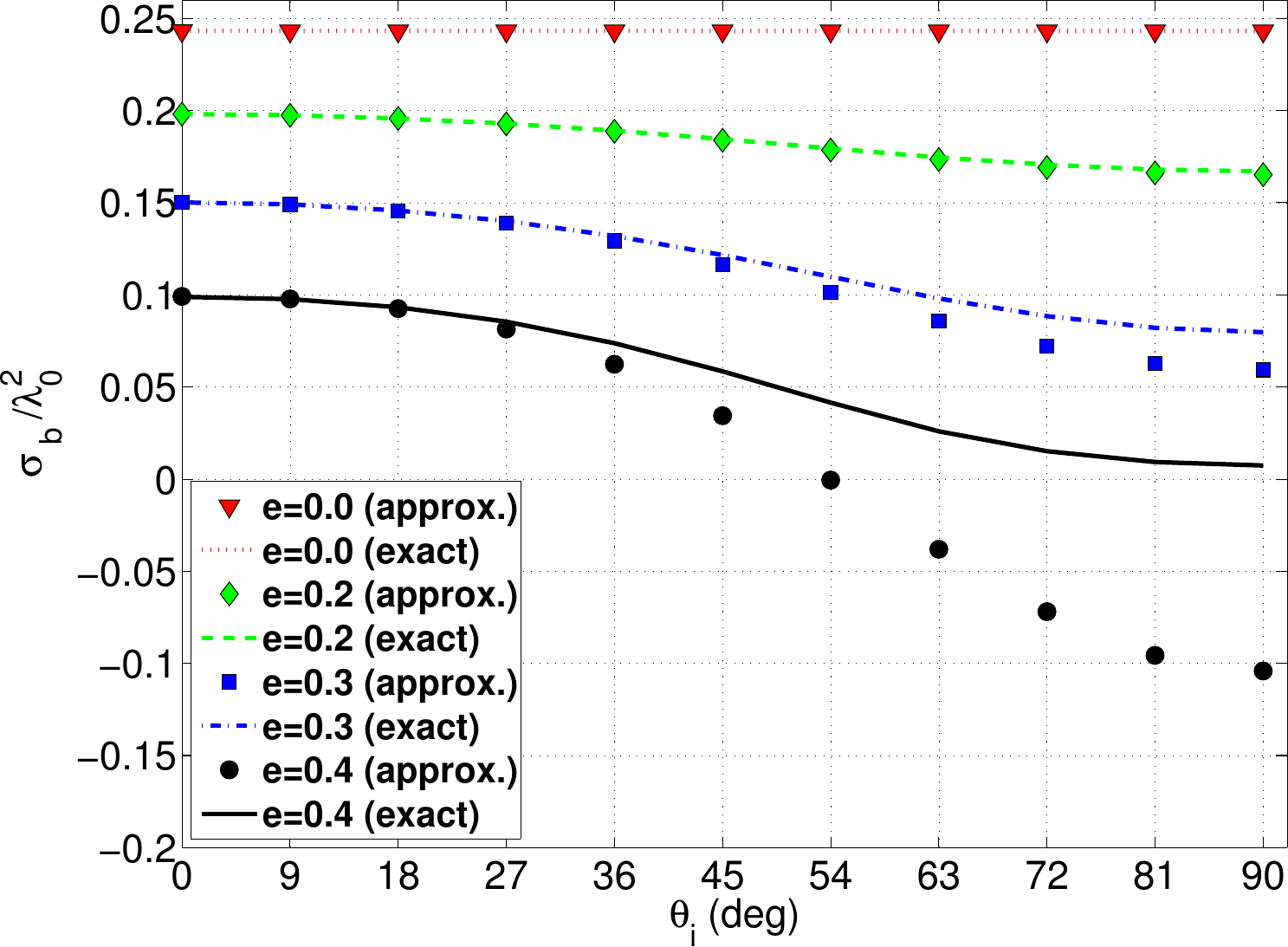} 
	\includegraphics[width=0.46\textwidth]{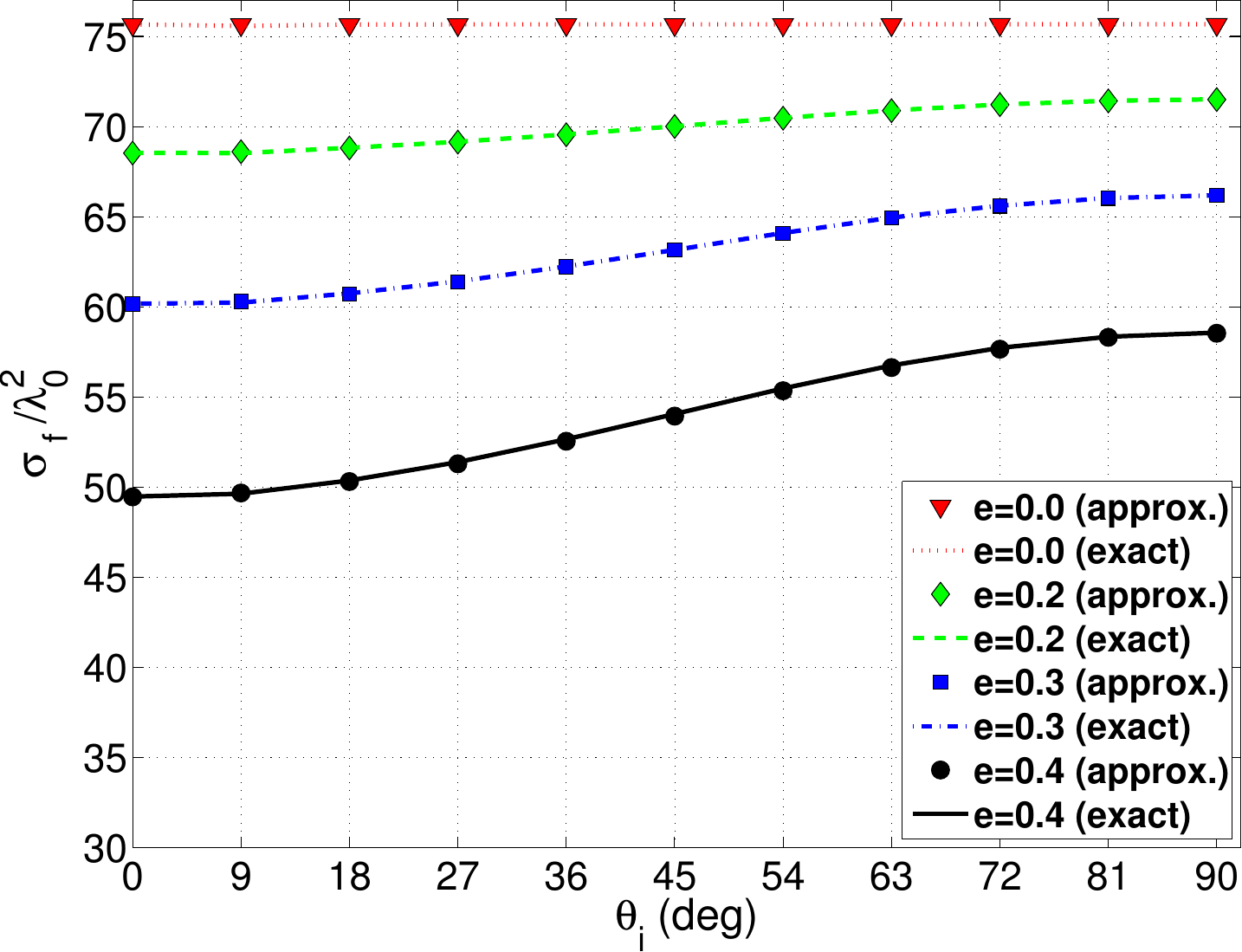}
	\caption{Comparison between exact solution and analytical approximation previously published for prolate 
	spheroids with different kinds of excentricity $\mathrm{e}$. Normalized back-scattering cross section (left) and
	forward-scattering cross section (right). Both are normalized by $\lambda_0$, the wavelength in 
	the medium.}
	\label{fig:KotsisRume}
\end{figure*}

\subsection{Comparison with reported results}

Among the relatively great number of publications addressing the acoustic scattering by penetrable spheroids, 
such as fluid spheroids immersed in a fluid surrounding medium, there is an article \cite{Kotsis} that reports 
an analytical solution of the scattered pressure field for $\rho_1/\rho_0 \not\approx 1$, 
$k_1/k_0\not\approx 1$ and small values of eccentricity $e$ (i.e. $e=d/(2a)\ll1$). In that paper a shape 
perturbation method is used to express the scattered pressure in terms of spherical wave functions, instead of 
spheroidal ones. Hence, approximate expressions are presented for any value of $e$ as a product of the 
scattering pressure by a sphere and a polynomial in even powers of $e$ till $\mathcal{O}(e^6)$. Moreover, in 
that reference some calculated expansion coefficients are presented in a Table for $\rho_1/\rho_0 =1.22$, 
$k_0/k_1 = 1.27$ and $a/\lambda_0=0.9$, being $\lambda_0$ the acoustic wavelength in the medium. Those 
computed values lead to forward and backscattering cross-sections, $\sigma_f$ and $\sigma_b$, plotted in 
Figures 5 and 6 of the above mentioned work that can be compared with the corresponding results of the exact solution 
for $|f_\infty|$, obtained through the codes implemented here for a penetrable prolate spheroidal scatterer, 
given the relationships
\[
\begin{aligned}
	\sigma_f &= 4\pi|f_\infty(\theta_s = \theta_{i}, \varphi_s = \varphi_{i})|^2  \\ 
	\sigma_b &= 4\pi| f_\infty(\theta_s = \pi-\theta_{i},\varphi_s = \varphi_{i}+\pi)|^2
\label{equivalenciasKotsis}
\end{aligned}
\]
in far-field conditions.

Comparison between normalized backscattering and forward scattering cross-sections (i.e. 
$\sigma/\lambda_0^2$) predicted by the approximate solution \cite{Kotsis} and
the exact one computed through the implementation presented here is shown in Figure \ref{fig:KotsisRume}. 
Good agreement is achieved for the forward scattering case while the agreement gets worse for increasing 
eccentricity when backscattering is considered. For $e=0.4$ (i.e. $ b/a > 0.9$, which is quite close to be a 
spherical shape), $\sigma_b$ takes undesirable negative values at incident angles greater than $54^\circ$, 
which has no physical sense whereas the exact solution predicts positive values as expected.

\section{Numerical results}
\label{applications}

\subsection{Applications to aquatic ecosystem research}

In acoustical oceanography, the Prolate Spheroidal Model (PSM) \cite{Spence, Furusawa,Prarioetal},  provides a 
useful tool for determining target strength (\textit{TS}) of individual fish and volume scattering strength 
$(S{_{V}})$ of zoo and phytoplankton. In this research field, it is useful to estimate their numerical 
abundance as well as to improve morphological and anatomical representations of the volume scatterers present 
at sea.

In fisheries acoustics, as well as in other SONAR applications, sound scattering by an object is analyzed in 
the logarithmic scale using the target strength parameter, $TS$, that
can be expressed as 
\ben
	TS_{bs}\:(\mbox{re. 1 m}) = 20 \log( |f_\infty(\theta_s = \pi-\theta_{i},\varphi_s = 
\varphi_{i}+\pi)| ),
\een 

Equation \eqref{finf}, has length dimensions. 
where $|f_\infty|$ has length dimensions. In particular, for the prolate 
spheroidal scatterer, $f_\infty$ is defined in Equation \eqref{finf}. 

Previously, other investigators using the PSM have reported solutions on scattering by fluid spheroids for 
some cases of interest such as the so-called \textit{weakly scattering} (i.e.  $c_1/c_0\approx 1$ and 
$\rho_1/\rho_0\approx1$) and the \textit{gas-filled} case  \cite{ye1998low}. More precisely, Furusawa 
\cite{Furusawa} worked with $1.01\leq\rho_1/\rho_0\leq1.07$. Other authors limited their calculations to 
certain range of frequencies \cite{jechetal2015, okumura2003acoustic}.

\subsection{Weakly-Scattering by a prolate fluid spheroid}

When the medium and the prolate spheroid have similar physical properties, i.e. $\rho_1 / \rho_0 \approx 1$  
and $c_1/c_0 \approx 1$, a usual approximation consists in assuming that the non-diagonal matrix elements 
$\alpha_{n\ell}^{(m)}$ defined in Equation \eqref{alpha} can be neglected \cite{Furusawa, Prarioetal, 
jechetal2015}, thus
\begin{equation} 
	\alpha_{n\ell}^{(m)} \approx \alpha_{nn}^{(m)} \delta_{n\ell},
	\label{aproximation2}
\end{equation}
whereas the matrix system given by Equation (\ref{matricial_discreto}) becomes almost trivial so that 
straightforward calculations lead to the approximate coefficients,  
\begin{equation}
	A^{\mbox{\tiny approx}}_{mn} = - \frac{E^{m(1)}_{n}}{E^{m(3)}_{n}}
	\label{approximation} 
\end{equation} 
where
\ben
	E_n^{m(i)}= R_{mn}^{(i)}\argg{h_0}{\xi_0} - \frac{\rho_1R_{mn}^{(1)}\argg{h_1}{\xi_0}}{\rho_0 
	{R}_{mn}^{(1)'}\argg{h_1}{\xi_0}} {R}_{mn}^{(i)'}\argg{h_0}{\xi_0}.
\een

It is worthy to note that the approximation indicated in Equation \eqref{aproximation2} leads to an explicit expression
for the $A_{mn}$ coefficients as set in Equation \eqref{approximation}. This crucial approximation, which avoids the 
numerically heavy task of solving the successive matrix systems, is very often used and it turns out to be appropriate in several cases of interest 
such as weakly scattering, low eccentricity and low frequencies. However, there are cases of interest where this approximation can not be used.
On the other hand, solving the exact case through the matrix system is somewhat tricky 
because the matrix $Q_{\ell j}^{(m)}$ may be ill-conditioned when $|c_1/c_0 - 1| \gg 1$.
To avoid errors emerging from the numerical solution of the matrix system, algebra of 
precision beyond hardware floating point (64 bits) has been employed.

In order to compare results in the case of weakly scattering, the value of TS derived from the exact 
coefficients' computation through the solution (Equation \eqref{matricial_discreto}) with the one derived from 
the 
approximate solution (Equation \eqref{approximation}), a prolate spheroid insonified at 38 kHz, has been 
considered. The input data have been $\rho_1=1028.9$ kg m$^{-3}$ and $c_1=1480.3$ m s$^{-1}$, $a=0.1$ m, 
$b=0.01$ m for the spheroid immersed in seawater characterised by $\rho_0=1026.8$ kg m$^{-3}$ and $c_0=1477.4$ 
m s$^{-1}$. 
		
\begin{figure}[ht]
	\centering
	 \includegraphics[width=0.45\textwidth]{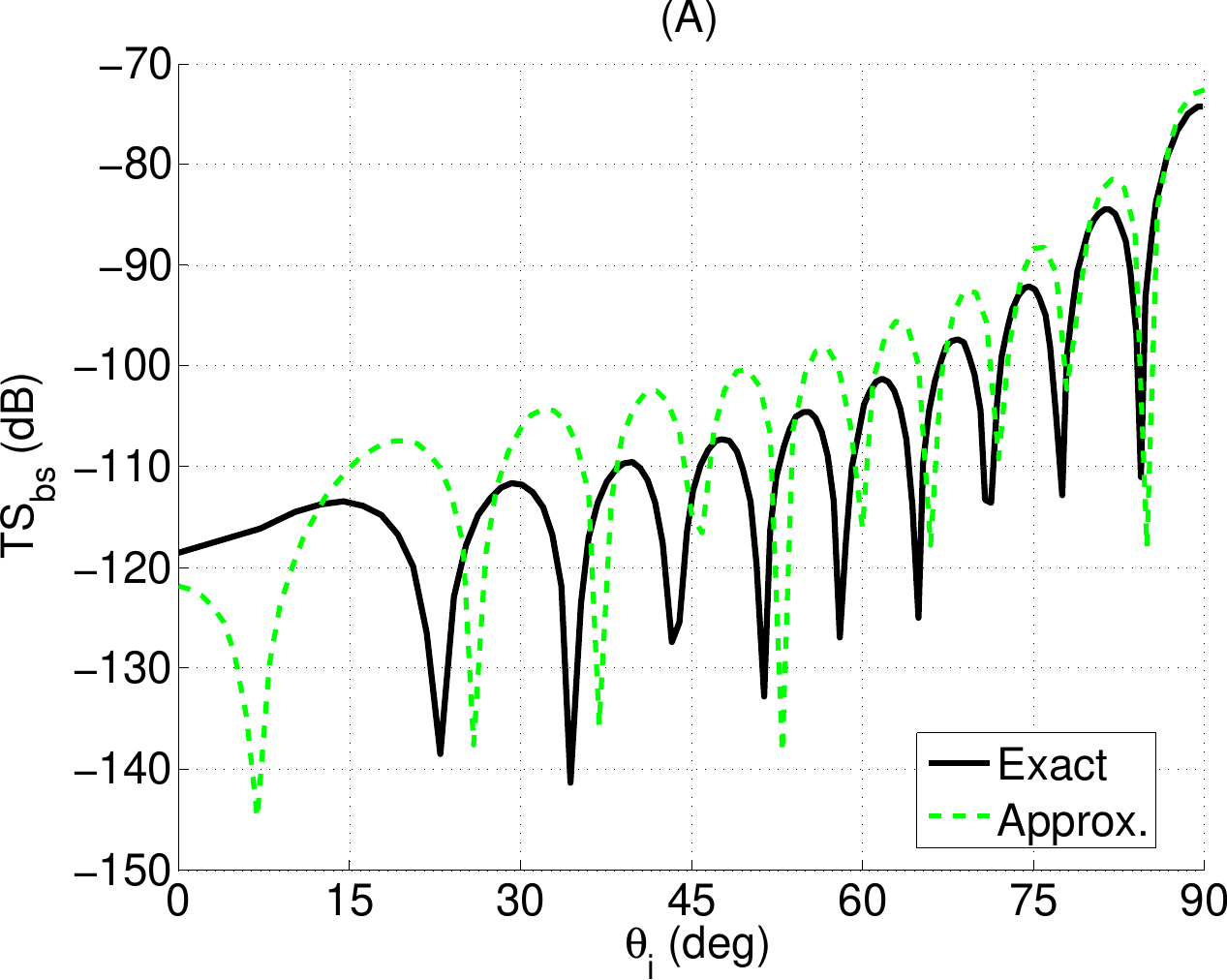} 
		\includegraphics[width=0.45\textwidth]{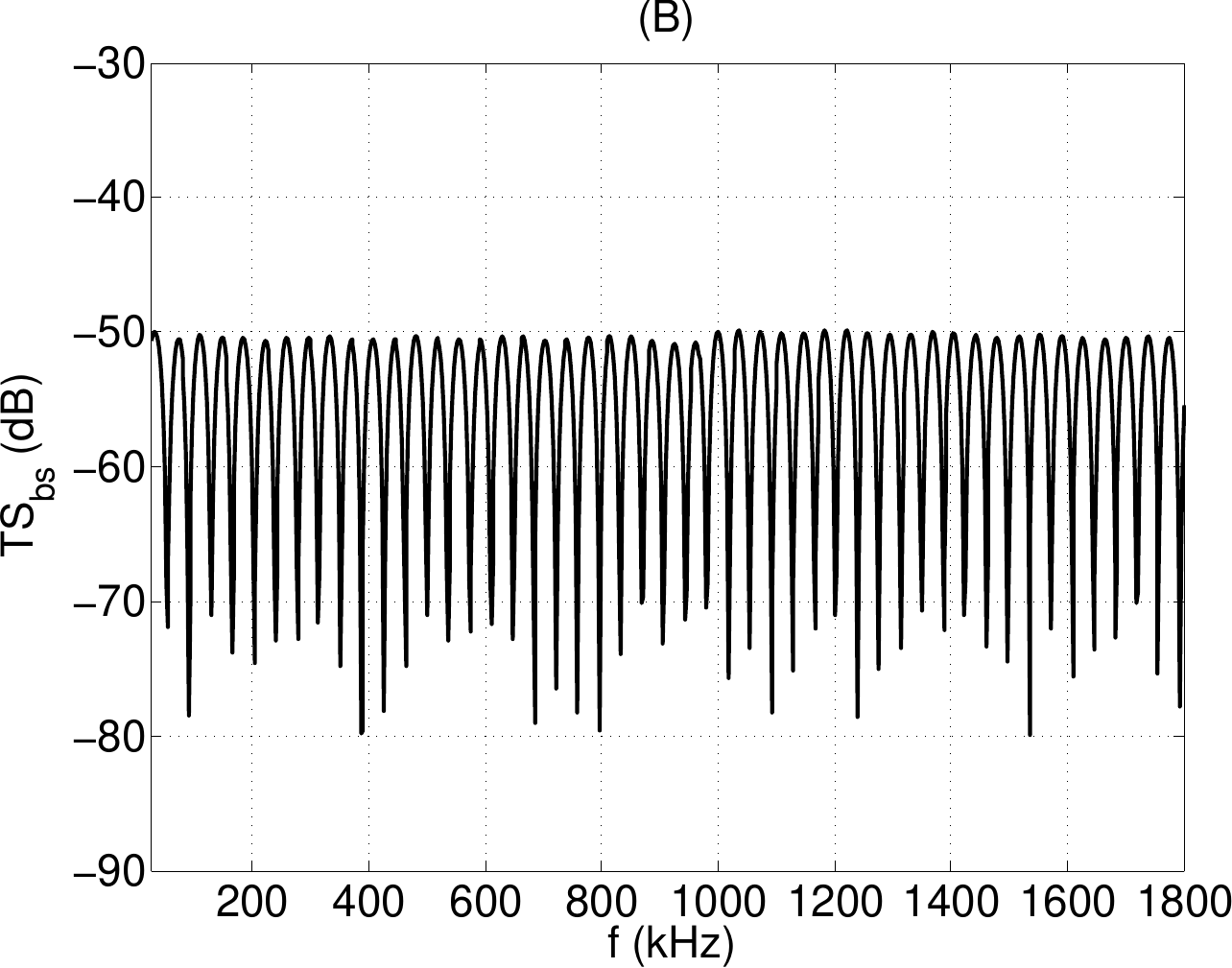}
	\caption{(A) TS \textit{vs.} incident angle, comparison between exact and approximate solutions for a prolate spheroid. $\rho_0=1026.8$kg m$^{-3} $, $\rho_1=1028.9$kg m$^{-3}$, $c_0=1477.4$m s$^{-1}$, $c_1=1480.3$m s$^{-1}$, $a=0.1$m, $b=0.01$m, $f=38$kHz. (B) TS \textit{vs.} frequency at $\theta_i=90^\circ$ for a spheroid with the same parameters than (A) but $\rho_1=1088.4$ kg m$^{-3}$.}
	\label{fig_Weakly}
\end{figure}

Comparison between both solutions is shown in Figure \ref{fig_Weakly}(A) for the whole $0^\circ$ to 
$90^\circ$ incidence angle-range. It can be observed that both curves do not fit. 
Furthermore, it should be noted that the
used approximation (Equation \eqref{aproximation2}) to compute the matrix elements $Q$, defined in Equation 
\eqref{QQ}, leads to detect nearly no differences between the exact and the approximate solutions for TS,
whereas using this approximation to compute the vector components $f$ (Equation 
\eqref{ff}), generates very significant differences when the density contrast verifies $\rho_1/\rho_0\lesssim 
1.002$ like in this case. However, if the sound speed contrast is $c_1/c_0\approx 1$ but the density contrast 
is for example $\rho_1/\rho_0=1.06 $, the differences between the approximate and exact solutions is roughly 
$0.6$ dB. 

Additionally, in the frame of the approximate solution to compute TS (Equation \eqref{approximation}), the 
maximum frequency range previously reported \cite{jechetal2015} could be extended till $h_0=762.08$. Results 
are shown in Figure \ref{fig_Weakly}(B) for 25 kHz $\leq$ f $\leq$ 1800 kHz and $\theta_i = 90^\circ$, where the series in Equation \eqref{pscatt} was truncated at $m \leq M=150$ and $n \leq M+m$, so 
that a total of 22801 modes were summed.
A previously published algorithm suitable for high values of $h$ was used in order to compute prolate 
spheroidal wave functions \cite{VanBurenSite}.

%
%

\subsection{Gas-Filled Prolate spheroid}

A gas-filled spheroid is a fluid spheroid with density and sound speed contrasts similar to the 
ones corresponding to the water-air interface (i.e. $c_1/c_0<1$ and $\rho_1/\rho_0\ll 1$). This case has a 
wide spectrum of applications in acoustical oceanography. In particular, at intermediate frequencies, the fish 
swimbladder may be considered as a gas-filled or a soft-scatterer when the PSM is used.   
\begin{figure}[ht]
	\centering
	 \includegraphics[width=0.75\textwidth]{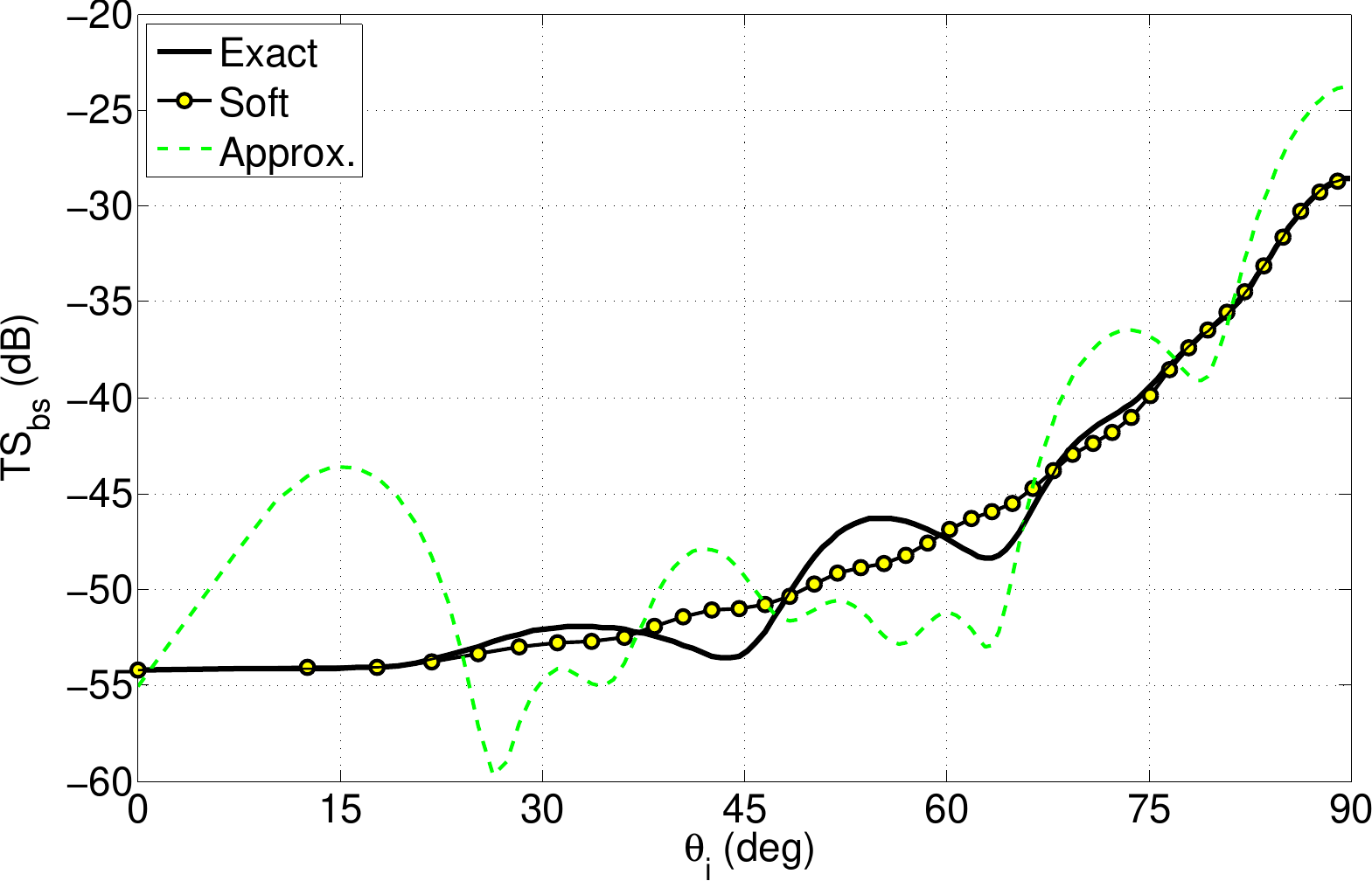} 
	\caption{\small{TS \textit{vs}. incident angle for a gas-filled.$\rho_0=1026.8$ kg m$^{-3} $, $\rho_1=1.24$kg m$^{-3}$, $c_0=1477.4$ m s$^{-1}$, $c_1=345.0$ m s$^{-1}$, $a=0.07$ m, $b=0.01$ m and f=38 kHz.}}
	\label{fig_GasFilled}
\end{figure}

At low frequency ($k_0 a \ll 1$), it is possible to use an approximation similar to Equation \eqref{aproximation2} 
instead of solving the exact Equation \eqref{matricial_discreto} \cite{yeh1964}. Scattering by a gas-filled prolate 
spheroid for $k_0 a \ll 1$ has already been theoretically investigated under the approach of the PSM, and 
some numerical results were published \cite{ye1998low}. However, at intermediate and high frequencies, the 
gas-filled prolate spheroid has not been modelled with the PSM \cite{jechetal2015, okumura2003acoustic}, 
hence, other approximate alternatives were used.
More precisely, Jech \textit{et al.} 
\cite{jechetal2015} reported that convergence was not achieved for 
the following input data: $\rho_0=1026.8$ kg m$^{-3}$, $\rho_1=1.24$ kg m$^{-3}$, $c_0=1477.4$ m s$^{-1}$, 
$c_1=345.0$ m s$^{-1}$, $a=0.07$ m, $b=0.01$ m and $f=38$ kHz. Then, in that work, TS values from a 
gas-filled spheroid were computed through different methods such as the Boundary Element Method and 
Kirchhoff ray mode. The same spheroid has been considered in this work, 
and convergence was achieved. Results are shown in  Figure \ref{fig_GasFilled}, where it can be seen 
that the soft-scatterer and the exact solution have the same behaviour. Furthermore, it can be observed 
that the approximate solution given by Equation \eqref{approximation} clearly cannot be used in this case. 
			
\section{Routines' sintaxis}
\label{sintaxis}

The provided routines for far-field patterns computation when plane acoustic waves incide on prolate or oblate spheroids 
were developed in Julia programming language, which has a MATLAB-like sintaxis, so that MATLAB and Octave 
users are expected to get easily familiarized with it. They can be freely downloaded from a GitHub repository \texttt{https://github.com/elavia/liquid\textunderscore spheroid} as a 
ZIP file or as individual files and they can be directly executed from the Julia REPL interface.

The material uploaded in the repository is organized in two folders, namely, Linux and Windows, since the routines have been 
implemented for both operative systems. They were developed and tested in Julia 0.5 under Fedora Linux and in Julia 
0.4 under Windows 7, respectively. For ensuring their correct functionality, previous installation of Julia's GSL package (Julia implementation of 
the GNU Scientific Library) is required\footnote{In most cases it is enough to type \texttt{Pkg.add(``GSL'')} 
at the Julia's prompt.}.

The files \texttt{JUL.routines.jl} and \texttt{JUL.auxiliar.jl} contain the main and auxiliary functions for 
the expansion coefficients and the far-field patterns calculations. 
Several Julia {\it scripts} reproducing the results presented in Section \ref{verifarfield} are also provided. 
Their names were chosen so that they are related to their particular purpose. Hence, it is not a surprise for 
example that \texttt{script.oblate\textunderscore liquid.jl} provides the pattern for an oblate spheroid in 
the fluid case.

The AGD executable files \texttt{obl\textunderscore sphwv} and \texttt{pro\textunderscore sphwv} constitute 
the {\it numerical engine} for the calculation of the spheroidal wave functions inside Julia routines and they are called 
through specialized shell scripts (\texttt{.sh} files, under Linux) and batch files (\texttt{.bat} files, 
under Windows). Those executable files were taken from the \texttt{scattering-master/spheroidal/sphwv} directory in the AGD 
software source tree and they can be downloaded from the authors' site at \texttt{https://github.com/radelman/scattering}. 

For a detailed description of the AGD software and the subtleties in the calculation of spheroidal wave functions, 
the reader is referred to the original reference that partially inspired this work \cite{Agd2014}. The 
parameters configuration for the AGD software execution inside the Julia routines has been done trough the 
text files \texttt{obl.parameters} and \texttt{pro.parameters}. In these files safe values of the parameters 
were taken in order to ensure reliable results.
It is remarkable that input selection associated with extreme cases, such as high frequency, may lead to 
an unpredicted behaviour, whose comprehension requires deeper understanding of the AGD code. However, these 
cases are far beyond the scope of the provided routines.

The first step to begin working with the code, once the Julia environment is properly installed, is to download 
the routines from the authors' repository, decompress them in a directory and start the Julia REPL interface
from that directory. Then, main and auxiliary functions must be loaded at the Julia's prompt through the script \texttt{JUL.main.jl},
before any other script is run. Afterwards all the routines are available in the current workspace and can be called 
from the prompt.

In order to illustrate and clarify the routines' operation, an example session is presented.
First, the routines are loaded through the {\it inclusion} of the main script, i.e.

\begin{verbatim}
	julia > include("JUL.main.jl") ;
\end{verbatim}

Therefore, the routines are in the workspace and it is now possible, for example, to calculate the far-field pattern $|f_\infty|$ in the 
fluid case for a prolate spheroid. This is accomplished with the command

\begin{verbatim}
	julia > include("script.prolate_liquid.jl") ;
\end{verbatim}

The pattern is saved in a file called \texttt{Out.Pattern.dat} within the current directory.
The data format in this file is: angle (in radians) and absolute value of the $f_\infty$ (in arbitrary length 
units compatible with the ones used for $a, b, k$).
The computing elapsed time is strongly dependent on the computer's speed and the parameters set in the AGD software.
All the provided scripts save the results in the same file so that the content is overwritten each time.
Scripts output can be plotted with any plotting software (In particular, the plots in this work were made using MATLAB and Gnuplot).

The scripts have a section of user-defined parameters.
In the above mentioned example, \texttt{script.prolate\textunderscore liquid.jl} contains

\begin{verbatim}
	# ~~~~~~~~~~~~~~~~~~~~~~~~~~~~~~~~~~~~~~~
	# User configurable parameters
	# ~~~~~~~~~~~~~~~~~~~~~~~~~~~~~~~~~~~~~~~
	
	# Physical parameters
	rho10 = 5.00 ; # Density ratio
	k_0 = 4 ; # Wave number in media 0
	k_1 = 6 ; # Wave number in media 1
	a = 1 ; # Major semiaxis spheroid
	b = 0.99 ; # Minor semiaxis spheroid
	theta_inc = pi/4 ; # Incidence angle (rad)
	
	# Software parameters
	M = 8 ;
	method = 2 ;
	delta_eta = 0.005 ;
\end{verbatim}

The meaning of the physical parameters, namely, $\rho_{10}, k_0, k_1, a, b, \theta_i$ is evident from the
explicative text after the commentary character \verb+#+.
The software parameters section includes parameters related to the spheroidal wave calculation.
The \verb+M+ parameter is the maximum value taken by $m$ in the coefficients $A_{mn}, B_{mn}$ and is
directly associated to the total number of the coefficients to be calculated.
For instance, \verb+M+$\:=8$ value means a total of $(M+1)(M+2)/2=45$ coefficients.
The \verb+method+ parameter refers to the type of calculation (\verb+method+$\:=1$ is convenient for $\xi \gg 1$ and 
\verb+method+$\:=2$ for $\xi \sim 1$).
Finally, \verb+delta_eta+ is the step on the $\eta$-grid and as a consequence of the inequality $-1 \leq \eta \leq 1$, that value leads 
to a $2/$\verb+delta_eta+ grid size.

\section{Discussion and conclusions}
\label{conclusions}

A novel implementation of the exact analytical solution for the problem of acoustic scattering by fluid or penetrable 
prolate/oblate spheroids governed by Helmholtz equation, is presented. A set of computational routines to 
calculate the expansion coefficients appearing in the mathematical expressions of the scattered and 
transmitted acoustic pressures is provided. These routines are freely available from the authors' GitHub 
repository. The coefficients' computation is precisely the most cumbersome task related to numerical 
calculation of the mathematical expressions above mentioned.

The results predicted by the computational implementation were tested for the geometrical limit when both 
prolate/oblate spheroids tend to the sphere and for the physical limit when they tend to soft or rigid 
spheroidal scatterers, taking low and high values of dentisities ratios, respectively. In the latter two 
cases predicted results turn out to be in agreeement with expected ones for Dirichlet and Neumann boundary 
conditions. 
Additionally, it has been qualitatively verified, trough the calculation for near-field conditions, that the 
numerical solution satisfies the boundary condition at the surface of the spheroids. The implemented exact 
solution has been successfully compared against a previously reported approximate solution and extended to 
eccentricity ranges where the approximation does not work properly.

Furthermore, the implemented solution enabled to extend the applicability ranges previously published in the 
literature for cases of interest in aquatic ecosystem research.
Table \ref{tabla1} summarizes the physical properties, the geometry and type of spheroid and the acoustic 
frequency range of the penetrable scatterer considered by different authors.
\begin{table}
\resizebox{0.8\textwidth}{!}{
 	\begin{tabular}{ |l|  l l  c  c c  |}
		\hline 
	&  &   &  &  & \\		
	\textbf{Reference} & $\rho_1/ \rho_0$ & $c_1/ c_0$  & $h_{0} (\text{max})$   & $b/a (\text{min})$ & 
Type of \\  
	&  &   &  &  & Spheroid \\ \hline
	&  &   &  & &  \\   	
	Yeh, 1967 \cite{yeh1967}& $0.50000$ & $ 1.000 $  & 9.38 & 0.109 & Prolate \\ 
	&  &   &  &  & \\		\hline 
	&  &   &  & &  \\   	
     Furusawa, 1988 \cite{Furusawa}& $1.07000$ & $ 1.050 $ & $12.00$  & $0.100$  & Prolate \\ 
	&  &   &  & &  \\   	\hline
	&  &   &  & &  \\   	
	Ye \textit{et al.}, 1998 \cite{ye1998low}& $0.00129$ & $ 0.220 $  & $0.73$   & $0.050$ & Prolate \\
	&  &   &  & &  (low freq.)\\   	\hline
	&  &   &  & &  \\   	
	Okumura \textit{et al.}, 2003 \cite{okumura2003acoustic}& $1.05000$ & $ 1.050 $  & $9.79$   & $0.200$ & Prolate \\
{} &  &   &  &  & \\   	\hline
	&  &   &  & &  \\   		
	Tang \textit{et al.},2009 \cite{Tang_Nishimori_Furusawa2009} & $1.04000$ & $1.020$  & $24.84$ & $0.150$ & 
Prolate\\
			&  &   &  &  & \\   	\hline
	&  &   &  & &  \\   
	Prario \textit{et al.}, 2015 \cite{Prarioetal} & $1.06800$ & $ 1.088 $  & $120.00$  & $0.238$ & Prolate\\ 
	&  &   &  &  & \\   	\hline
	&  &   &  & &  \\   			
	Jech \textit{et al.}, 2015 \cite{jechetal2015}& $1.00200$ & $ 1.002 $  & $118.00$  & $0.143$ & Prolate \\  
	&  &   &  &  & \\   	\hline
	&  &   &  & &  \\   	
	& $3.00000 $ & $ 0.600 $  & $762.08$  & $0.100$ & \\
	This work{} & $0.00100 $ & $ 0.220 $ &  & &Prolate/ \\   
	{} & $1000.00000 $ & $1.500 $  &   &  &Oblate\phantom{ll}\\     
	{} & $0.00129 $ & $1.270 $  &   &  &\\     	\hline 
		\end{tabular}
	} 
	\vspace*{2mm}
		\caption{ Physical properties, the geometry and type of spheroid and the 
		acoustic frequency range of the penetrable scatterer considered by different authors (i.e. 
		density and sound speed contrasts; $h_0$ maximum, $b/a$ ratio and prolate and/or oblate 
		spheroid).
	\label{tabla1}
}
\end{table}

With the aim of generating the computational codes presented in this work, the efficient software by AGD was used in 
addition to the numerical capabilities of the new scientific programming language Julia. Thus, it was possible 
to take advantage of the arbitrary floating point precision support in both pieces of code. Moreover, 
scripts to reproduce the calculations presented in Section $4.1$ are also provided to enable a minimum 
consistency check. 

The developed codes are released intending to contribute with researchers working in acoustic scattering in order to
simplify the tedious algorithmic task involved in the resolution of the cumbersome matrix system when 
calculating the exact solution for a fluid spheroidal scatterer. 
\section*{\textbf{Acknowledgements}}

This work was supported by the PIDDEF Program of the Argentinian Ministry of Defense (13-2014), the
Argentinian Navy and the National Council for Scientific and Technical Research (CONICET).

The authors wish to remark once more the outstanding piece of software that Adelman and his collaborators have made freely available to the scientific
community. It was a real pleasure to work with it. Of course, the same applies to Julia programming language. Moreover, they acknowledge the work of the programmers' team 
that have designed and implemented that language. It is marvelous. With both they are in debt. 

\end{document}